\documentclass[aps,prd,onecolumn,superscriptaddress,showpacs,floatfix,10pt]{revtex4}
\usepackage{amsfonts}
\usepackage{subfigure}
\usepackage{graphicx}
\usepackage{epsfig}
\usepackage{amsmath}
\usepackage{slashed}
\usepackage{units}

\begin{document}

\title{The fundamental problem of treating light incoherence in photovoltaics and its practical consequences}

\author{Aline Herman$^{\dagger}$}
\email{aline.herman@unamur.be}
\author{Micha\"{e}l Sarrazin$^{\dagger}$}
\email{michael.sarrazin@unamur.be}
\author{Olivier Deparis$^{\dagger}$}
\email{olivier.deparis@unamur.be}

\affiliation{Solid-State Physics Laboratory, Research Center in Physics of Matter and Radiation (PMR), University of Namur, 61 rue de Bruxelles, B-5000 Namur, Belgium\\$^{\dagger}$ These authors have contributed equally to this work.}

\begin{abstract}
The incoherence of sunlight has long been suspected to have an impact on solar cell energy conversion efficiency, although the extent of this is unclear. Existing computational methods used to optimize solar cell efficiency under incoherent light are based on multiple time-consuming runs and statistical averaging. These indirect methods show limitations related to the complexity of the solar cell structure. As a consequence, complex corrugated cells, which exploit light trapping for enhancing the efficiency, have not yet been accessible for optimization under incoherent light. To overcome this bottleneck, we developed an original direct method which has the key advantage that the treatment of incoherence can be totally decoupled from the complexity of the cell. As an illustration, surface-corrugated GaAs and c-Si thin-films are considered. The spectrally integrated absorption in these devices is found to depend strongly on the degree of light coherence and, accordingly, the maximum achievable photocurrent can be higher under incoherent light than under coherent light. These results show the importance of taking into account sunlight incoherence in solar cell optimization and point out the ability of our direct method in dealing with complex solar cells structures.
\end{abstract}


\maketitle

\section{Introduction}

In photovoltaics, it has long been suspected that the incoherence of
sunlight has an impact on the energy conversion efficiency of solar cells, although the extent of this is unclear. In
photosynthesis, on the other hand, the incoherence of sunlight has
been recently recognized to play a fundamental role on the
optical-biological energy conversion process \cite{Mancal,Kassal}. 

The use of ultrathin (a few microns) crystalline Silicon (c-Si) active layers in solar cells is promising since it requires less material and therefore to decrease the costs. However, ultrathin layers result in drastic reduction of the absorption of the solar radiation in the near-infrared region due to the indirect band-gap of c-Si \cite{Muller}. The efficiency of ultrathin solar cells is therefore limited. The use of optimized periodic photonic nanostructures (light-trapping structures) at the front and$\backslash$or back-side(s) of the active layer of the solar cell is a promising approach to solve this issue \cite{Tsakalakos,Nelson,Zeman,Campbell,Yablonovitch,Abass}.
This well-known design helps to couple incident light into the active layer via quasi guided modes \cite{Zeman,Campbell,Yablonovitch,Saeta,Gomard,Yu_PRL}.
Most research focuses on finding the optimal structure geometry \cite {Sigmund,Gjessing_2010,Bozzola,Herman,Jovanov,Lockau} that increases the absorption inside surface-corrugated ultrathin layers with the aim of reaching the fundamental upper bound limit on absorption \cite{Yu_PRL,Yu_OE,Niv,Markvart,Mellor,Naqavi,Naqavi_2013}. The determination of optimal light-trapping structures in ultrathin solar cells is therefore of high interest without loss of generality. This is the reason why we choose to focus here on the particularly relevant case of an ultrathin c-Si slab having its front side corrugated with periodic nanostructures (figure \ref{fig1}).  
Nevertheless, at present, the important issue of the plausible impact
of sunlight incoherence on cell efficiency remains quite unexplored.
Indeed, it is well known that the response of optical devices
depends on the degree of coherence of the incident light
\cite{BornWolf}. Until now, the rarity of investigations in this
area was related to the complexity of numerical methods dealing with
incoherence. Methods addressing both spatial
\cite{Mitsas,Prentice_99,Prentice_2000,Katsidis,Centurioni,Troparevsky,KRC,Santbergen,Abass_JAP}
or temporal \cite{Lee} incoherence exist. The problem of incoherence
seems, in principle, theoretically resolved. However, apart from experimental optimizations \cite{Sai,Sai_2013}, the theoretical optimization of complex solar cells (corrugated
multilayers) under incoherent light has never been performed. This
bottleneck is due to practical limitations of computational methods
used to deal with incoherence. Simply stated, at each
wavelength, multiple independent computational runs are performed
and then statistically treated \cite{Lee}. Each individual run
consists of the resolution of Maxwell's equations in complex
inhomogeneous media using Rigorous Coupled Wave Analysis (RCWA)
\cite{Moharam,Sarrazin_PRB,Vigneron,Vigneron_SPIE} or
Finite-Difference Time-Domain method (FDTD)\cite{FDTD1,FDTD2}. For
each run, the phase of the incident
wave is randomly chosen \cite{Lee}. Since the treatment
of each wavelength needs multiple runs, the
computational time demand is much more severe in the incoherent case
than in the coherent one (where only one run is needed).
Furthermore, as the complexity of the solar cell structure
increases, the time required to compute one run
increases dramatically. Therefore, because of both the complexity of the cell and the complexity of the algorithmic method, the accurate modelling of a complex solar cell under
incoherent light becomes a formidable task. This is probably the
reason why the effects of sunlight incoherence on complex solar
cell efficiency have never been properly investigated.

In a recent article, we
developed a rigorous theory accounting for the effects of temporal
incoherence of light on the response of solar cells \cite{Sarrazin}.
In the proposed method, a single time-consuming computation step is needed: the electromagnetic calculation of the coherent absorption spectrum. The incoherent absorption spectrum is then deduced directly through a convolution product with the coherent absorption spectrum. This second step is totally independent of the first one and therefore no multiple runs are needed at all.
Our method is not only simpler than previous ones \cite{Mitsas,Prentice_99,Prentice_2000,Katsidis,Centurioni,Troparevsky,KRC,Santbergen,Abass_JAP,Lee} but it also leads to a drastic reduction in computational time.
Since the incoherent treatment (second step) is totally independent of the complexity of the solar cell structure, our method paves the route for extensive optimizations of solar cells under incoherent illumination.

In the present article, we show
that the degree of sunlight coherence has a dramatical, unsuspected impact on
the way solar cells should be optimized. Especially, we predict that the
photocurrent produced by a corrugated thin-film solar cell strongly depends on the coherence time
of the incident light. As an illustration, the maximum achievable photocurrent is numerically calculated in two types
of uncoated corrugated semiconductor slabs, namely crystalline Silicon (c-Si) and Gallium Arsenide
(GaAs), under exposure to incoherent light . The
slabs have their top surfaces corrugated
with wavelength-scale arrays of square or cylindrical holes for light trapping purposes. Though a real solar cell comprises more layers than the corrugated active material layer (antireflection coating, back reflector, electrodes etc.), the stand-alone corrugated slab is sufficient to highlight the effect of sunlight incoherence as it is intended here.
The physical mechanism responsible for the dependence of the photocurrent on the degree of sunlight coherence is then discussed. Finally, we predict the potential gain in computational time thanks to the proposed method. 

\section{Overview of the method}

A solar cell, like any optical-electrical energy conversion device, is at the same time a linear optical system as well as a photo-detector. Indeed, the cell performs light harvesting and, as in any linear system, is characterized by its transfer function (optical absorption spectrum here). The cell, on the other hand, collects electrons and holes which are generated by harvested photons.
In comparison with the sunlight coherence time (estimated to 3 fs \cite{Hecht}), the detector response $T$ is slow since typical carrier life time ranges  from
0.1 ns to 1 ms in silicon, according to the doping level \cite{Seraphin}.
Therefore, it is crucial to consider the slowness of the detector response when averaging the solar cell response (photocurrent) under incoherent excitation.

The maximum achievable photocurrent $J$ supplied by a solar cell is
given by \cite{Henry}:
\begin{equation}
J=\frac e{hc}\int A(\lambda )S(\lambda )\lambda d\lambda =\int J(\lambda
)d\lambda   \label{current0}
\end{equation}
where $e$ is the electron charge, $h$ is the Planck's constant, $c$ is the light velocity, $A(\lambda)$ is the active layer absorption spectrum, $S(\lambda)$ is the
global power spectral density (PSD) of the solar radiation (AM1.5G spectrum) and $J(\lambda )$ is the maximum
achievable photocurrent spectrum. 
It should be noted that the only quantity that is detected by a solar cell is the integrated photocurrent $J$. Therefore, the numerical computation of the absorption spectrum is solely a computational step towards the determination of $J$.
In order to take into account the
incoherent nature of sunlight, $A(\lambda)$ must represent
the \textit{effective} incoherent absorption $A_{incoh}(\lambda)$ undergone
by the solar cell. By \textit{effective} absorption, we mean that $A_{incoh}(\lambda)$ has to be considered as an intermediate quantity for calculating $J$ (see later discussion related to figure \ref{fig2}). In numerous previously published works \cite{Gjessing_2010,Bozzola,Herman,Gomard,Gjessing}, $A(\lambda)$ is actually the coherent
absorption $A_{coh}(\lambda )$ which is computed using numerical methods (RCWA, FDTD)
that propagate the coherent electromagnetic field. In a few works,
numerical methods were proposed in order to compute $A_{incoh}(\lambda )$
\cite{Mitsas,Prentice_99,Prentice_2000,Katsidis,Centurioni,Troparevsky,KRC,Santbergen,Abass_JAP,Lee}. 
However, they rely on multiple numerical runs, each one being performed for a
coherent incident wave which is randomly dephased with respect to the previous one. The final
result is then obtained from statistical averaging. Though correct, this procedure is time-consuming and not necessary as we show it hereafter. 
Recently, we have shown that
$A_{incoh}(\omega)$, where $\omega =2\pi c/\lambda $ is the angular frequency, can be directly
obtained from the convolution product (noted $\star$) between $A_{coh}(\omega)$
and an incoherence function $I(\omega )$ \cite{Sarrazin}:
\begin{equation}
A_{incoh}(\omega )=I(\omega )\star A_{coh}(\omega ).  \label{F1}
\end{equation}
The incoherence function is
defined by the Gaussian distribution \cite{Sarrazin}:
\begin{equation}
I(\omega )=\tau _c\sqrt{\frac{\ln 2}{\pi ^3}}e^{-\frac{\ln 2}{\pi ^2}\tau
_c^2\omega ^2}  \label{gaussian}
\end{equation}
with a Full Width at Half Maximum $\Delta \omega =2\pi /\tau _c$ inversely
related to the coherence time $\tau _c$. Physically, $I(\omega )$ describes
the stochastic behaviour of each spectral line (optical carrier at frequency $\omega$)
composing the whole solar spectrum. This formula is easy to use in
practice and allows to reduce the algorithm complexity, hence the computational time. 
Full rigorous demonstration of (\ref{F1}) was
given in \cite{Sarrazin}. Nevertheless, in order to understand the physics
behind the convolution formula, we present a simplified version of the method reported in \cite{Sarrazin}.

\section{Theoretical framework of the method and physical interpretation}
Though  Maxwell's equations are linear, addressing the issue of the power flux absorbed by a linear system under incoherent excitation is not a trivial problem as we will see in the following sections. In the frame of random signal theory, we demonstrate hereafter that the incoherent output power of a linear system can be obtained from the coherent output power.
This general result applies to solar cells in particular, where the incident sunlight is temporally incoherent, i.e. each frequency component of the solar spectrum can be regarded as a random process. Since all random processes related to each optical carrier frequency are independent, each carrier frequency can be treated individually.

\subsection{Basic concepts in random signal theory}
Hereafter, we present briefly basic concepts in random signal theory such as autocorrelation,PSD and normalized power (we follow the notation of \cite{signal_processing}).
The real stationary random signal we consider is noted $x(t)$. In the particular case of solar cells, $x(t)$ is the electric field of the electromagnetic radiation.
The autocorrelation function of the random signal is defined as \cite{signal_processing}
\begin{equation}
R_{X}(\tau)=E[x(t)x(t+\tau)],
\label{autocor}
\end{equation}
where $E[]$ denotes the expectation value of $x(t)$ (i.e. ensemble average).
When $\tau=0$, we find the mean square value of the signal:
\begin{equation}
R_X(0)=E[x^2(t)].
\label{Power_coh}
\end{equation}
In the context of solar cells, this quantity is proportional to the average power transported by the optical wave at the carrier frequency $\omega_c$.
The Power Spectral Density (PSD) $S_X(\omega)$, is defined as the Fourier transform of $R_{X}(\tau)$:
\begin{equation}
S_{X}(\omega)=\int_{-\infty}^{\infty}{R_{X}(\tau) e^{i\omega \tau} d\tau}.
\end{equation}
However, for a stationary random signal expanding from $-\infty$ to $\infty$ in time, the function $R_X$ is not integrable \cite{signal_processing}.
Thus, the Fourier transform (hence PSD) does not converge.
In order to define the PSD of a random signal, the signal must be truncated within a span of time $T$, i.e. the sampling interval \cite{signal_processing}. The truncated signal is noted by $x_T(t)$, with  $x_T(t)=x(t)$ over time span $T$ and $x_T(t)=0$ elsewhere.
Thanks to truncation, the Fourier transform can be defined for each realization $x_T(t)$ of the signal. The stochastic quantity corresponding to the Fourier transform of the truncated signal is defined by:
\begin{equation}
X(\omega)=\mathcal{F}\left[x_T(t)\right],
\end{equation}
where $\mathcal{F}$ is the Fourier transform.
For large $T$, it can be shown \cite{signal_processing} that
\begin{equation}
S_{X}(\omega)=E\left[\frac{1}{T}|X(\omega)|^2\right].
\label{S}
\end{equation}
Using the PSD, we can then define the normalized average power as
\begin{equation}
P_X=R_X(0)=\frac{1}{2\pi}\int_{-\infty}^{\infty}{S_{X}(\omega) d\omega}=E\left[\frac{1}{T}\frac{1}{2\pi}\int_{-\infty}^{\infty}|X(\omega)|^2 d\omega \right].
\label{power}
\end{equation}
From a physical point of view, (\ref{power}) simply means that the integration of the PSD yields the power.

In the context of solar cells, the sampling time $T$ is effectively the photo-detector response time which is very long at the time scale of the random process.
Therefore, the assumption of large $T$ in (\ref{S}) is fully satisfied.
Note that, in the scattering matrix treatment of (\ref{F1}), the detector response was lumped in the time averaged Poynting vector flux expression in the form of a narrow bandwidth filtering function ((A26) to (A29) in \cite{Sarrazin}).

Hereafter, we consider a linear system subject to both coherent or incoherent input signals and we calculate the corresponding output signals.

\subsection{Coherent signal output}
The coherent input signal $x_{in}^{coh}(t)$ is taken to be a real cosine function:
\begin{equation}
x_{in}^{coh}=E_0 \cos(\omega_c t),
\end{equation}
where $E_0$ is the amplitude of the signal and $\omega_c$ is the carrier frequency.
According to (\ref{autocor}), we find
\begin{equation}
R_{x,in}^{coh}(\tau)=\frac{|E_0|^2}{2}\cos(\omega_c\tau).
\end{equation}
Since the coherent input power corresponds to $R_{x,in}^{coh}(\tau=0)$, we have the well known result:
\begin{equation}
P_{x,in}^{coh}=\frac{|E_0|^2}{2}.
\label{power_coh_in}
\end{equation}
According to linear system theory \cite{signal_processing} applied to deterministic signals, the coherent output power is given by:
\begin{equation}
P_{x,out}^{coh}=|G(\omega_c)|^2 P_{x,in}^{coh}=|G(\omega_c)|^2 \frac{|E_0|^2}{2},
\label{power_coh_out}
\end{equation}
where $G(\omega_c)$ is the transfer function of the system under study at frequency $\omega_c$.

\subsection{Incoherent signal output}
The incoherent input signal $x_{in}^{incoh}(t)$ is expressed as a carrier whose amplitude is randomly modulated:
\begin{equation}
x_{in}^{incoh}(t)=E_0 m(t) e^{-i\omega_c t},
\label{x_in_incoh}
\end{equation}
where $m(t)$ represents the random process (modulation function) and $e^{-i\omega_c t}$ is the periodic component of the signal at the carrier frequency.
The complex form of (\ref{x_in_incoh}) is simply used for the sake of simplicity in the mathematical derivation. Of course, we implicitly work with the real part of (\ref{x_in_incoh}).

The Fourier transform of the random input signal is
\begin{equation}
X_{in}^{incoh}(\omega)=E_0 M(\omega-\omega_c),
\label{X_incoh}
\end{equation}
where $M(\omega)$ is the Fourier transform of $m(t)$.
Using (\ref{power}) and (\ref{X_incoh}), we find for the incoherent input power:
\begin{equation}
P_{x,in}^{incoh}=|E_0|^2 E\left[ \frac{1}{T} \frac{1}{2\pi}\int_{-\infty}^{\infty}|M(\omega - \omega_c)|^2 d\omega\right].
\label{P_1_inco}
\end{equation}
According to (\ref{S}), we rewrite (\ref{P_1_inco}) as:
\begin{equation}
P_{x,in}^{incoh}=\frac{|E_0|^2}{2\pi}\int_{-\infty}^{\infty}S_M (\omega - \omega_c) d\omega,
\end{equation}
where
\begin{equation}
S_M (\omega)=E\left[\frac{1}{T} |M(\omega)|^2 \right],
\label{S_M}
\end{equation}
 is the PSD of the random process, an even function of $\omega$ \cite{signal_processing}.
Since both incoherent and coherent input signals must have the same power (i.e. $P^{coh}_{x,in}=P^{incoh}_{x,in}$),  $S_M(\omega)$ must verify the normalization relation:
\begin{equation}
\int_{-\infty}^{\infty}S_M (\omega) d\omega=\pi.
\label{normalization}
\end{equation}

Now, let us calculate the incoherent output power ($P_{x,out}^{incoh}$).
Using (\ref{power}), we get:
\begin{equation}
P_{x,out}^{incoh}=E\left[ \frac{1}{T} \frac{1}{2\pi}\int_{-\infty}^{\infty}|X_{out}^{incoh}(\omega)|^2 d\omega\right]
\label{P_out_inco_1}
\end{equation}
According to linear system theory, the Fourier transform of the output signal is related to the Fourier transform of the input signal through the transfer function \cite{signal_processing}:
\begin{equation}
X_{out}(\omega)=G(\omega) X_{in}(\omega).
\label{X_out}
\end{equation}
Then, it follows from (\ref{P_out_inco_1}) and (\ref{X_out}) that
\begin{equation}
P_{x,out}^{incoh}=E\left[ \frac{1}{T} \frac{1}{2\pi}\int_{-\infty}^{\infty}|G(\omega)|^2|X_{in}^{incoh}(\omega)|^2 d\omega\right].
\label{P_1}
\end{equation}
Using (\ref{X_incoh}) and (\ref{P_1}) we get:
\begin{eqnarray}
P_{x,out}^{incoh}& =&  \frac{|E_0|^2}{2\pi} E\left[ \frac{1}{T}\int_{-\infty}^{\infty}|G(\omega)|^2 |M(\omega - \omega_c)|^2 d\omega \right] \\
      &=& \frac{|E_0|^2}{2\pi} E\left[ \frac{1}{T}|G(\omega_c)|^2 \star |M(\omega_c)|^2 \right] \\
      &=& \frac{|E_0|^2}{2\pi} |G(\omega_c)|^2 \star E\left[ \frac{1}{T}  |M(\omega_c)|^2 \right].
\end{eqnarray}
The symbol $\star$ designates the convolution product.
Using (\ref{S_M}), we get:
\begin{equation}
P_{x,out}^{incoh}=\frac{|E_0|^2}{2\pi} |G(\omega_c)|^2 \star S_M(\omega_c).
\end{equation}
Considering the expression of the coherent output power (\ref{power_coh_out}), we find the relationship between the incoherent output power and the coherent one:
\begin{equation}
P_{x,out}^{incoh}(\omega_c)=\frac{1}{\pi} P_{x,out}^{coh}(\omega_c) \star S_M(\omega_c).
\end{equation}

Finally, using the normalization relation (\ref{normalization}), we find the most important formula of this paper:
\begin{equation}
P_{x,out}^{incoh}(\omega_c)=P_{x,out}^{coh}(\omega_c) \star I(\omega_c),
\label{P_final}
\end{equation}
where
\begin{equation}
I(\omega)=\frac{S_M(\omega)}{\int_{-\infty}^{\infty}S_M (\omega) d\omega}.
\end{equation}
As a consequence, the power of the incoherent output signal is equal to the power of the coherent output signal that is convoluted with the PSD of the random process.

Equation (\ref{P_final}) is formally equivalent to (\ref{F1}) which was derived in \cite{Sarrazin}. However, in the present case, we  have only considered the transfer function of a linear system relating a single input channel to a single output channel. This formalism obviously does not allow to calculate optical reflectance ($R$), transmittance ($T$) or absorption ($A=1-R-T$). In order to calculate these quantities, we must use the scattering matrix of the system instead of its transfer function. In this case, the derivation of (\ref{F1}) is more complicated (see \cite{Sarrazin} for details) but ends up with a formula equivalent to (\ref{P_final}).
It should be noted that a result similar to (\ref{P_final}) was reported for the description of vibrational random processes \cite{Mark}.
Surprisingly, albeit fundamental, the theory of linear system response to stochastic signals appears to have remained rather confidential until now, at least in the field of photovoltaics.

\section{Illustration of the method}
\begin{figure}
\centerline{\ \includegraphics[width=7.5 cm]{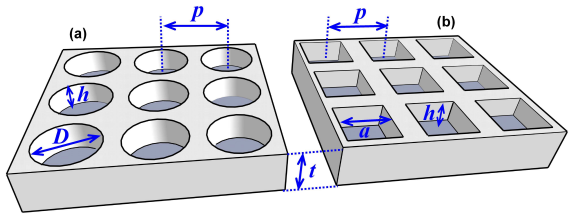}}
\caption{(Colour online). Corrugated slabs (i.e. active layers of
thin-film solar cells). $t$: slab thickness, $h$: height of
holes, $p$: period of hole array. $D$: diameter of cylindrical holes (a). $a$: side of square holes (b).}
\label{fig1}
\end{figure}

Using the above described theoretical method, we now investigate the effect of finite temporal coherence of the sunlight on
the efficiency of solar cells. Under coherent illumination, the type of front-side
corrugation is known to have a strong influence on the photocurrent \cite
{Bozzola,Herman,Gjessing}. Therefore, it is important to properly optimize
the corrugation (we do not consider here additional improvements brought by conformal antireflection coating and/or back reflector).
In order to investigate the impact of the coherence time on
such an optimization, we studied two different corrugations,
with cylindrical or square holes (figure \ref{fig1}).

Both corrugated slabs
have fixed thickness ($t=1\mu$m) and fixed hole depth ($h=500$nm). Slab thickness and hole depth are typical of ultrathin solar cell designs
where photonic light trapping effects are exploited \cite{Herman,Yu_PRL}.
The ratio between hole size (diameter $D$ or side $a$) and period ($p$) has a
fixed value equal to $0.9$. These values result from previous optimization studies \cite
{Herman}. The period is the only varying parameter
(from $250$nm to $1250$nm). Two types of materials are investigated:
crystalline silicon (c-Si) and gallium arsenide (GaAs). The holes are
supposed to be filled with air (i.e. incidence medium). The aim here is not to find again the best
corrugation shape (cf. \cite{Herman}) but to highlight the effect of coherence time for
different shapes.

The photocurrent (\ref{current0}) under incoherent illumination was
calculated for various coherence times (integration carried out from
$\lambda=200$nm to $2500$nm). Under incoherent illumination,
$A(\lambda)$ was equal to $A_{incoh}(\lambda)$ which was deduced from
$A_{coh}(\lambda)$ using (\ref{F1}). The coherent
absorption spectrum was numerically calculated using the RCWA method. 
In (\ref{current0}), the rate of incident photons (per unit area) at each carrier wavelength was fixed by the intensity of the corresponding frequency-resolved component of the AM1.5 solar power density spectrum. The incident light was
supposed to be unpolarized and impinging under normal incidence.

Before studying the photocurrent dependence on the coherence time, let us consider incoherent effects from a quantum mechanical point of view by resorting to Heisenberg's uncertainty principle. 
Due to their finite coherence time, each photon from the
solar radiation cannot be defined with a definite energy $E_0$ (or carrier wavelength
$\lambda _0=hc/E_0$). According to Heisenberg's uncertainty principle:
\begin{equation}
\Delta E\Delta t\geq \frac \hbar 2,  \label{Heisenberg}
\end{equation}
each photon is characterized by a spectral width $\Delta E$ with a time
uncertainty $\Delta t\approx \tau _c$ related to the coherence time, i.e. each photon is treated as a wave packet. Accordingly, each photon has finite coherence length, i.e. $l_c=c \tau_c$.
From (\ref{Heisenberg}), we must consider that a photon wave packet with a carrier
wavelength $\lambda _0$ occupies a spectral domain roughly defined by
$\mathcal{D}_{s} \sim \left[ \lambda_{0} - \Delta \lambda, \lambda_{0} + \Delta
\lambda \right] $ with $\Delta \lambda \approx \lambda _0^2/(4\pi c\tau _c)$%
. It means that the photon does not feel a single value of the complex refractive index $%
n(\lambda _0)+ik(\lambda _0)$, but a range of values $n(\lambda
)+ik(\lambda )$ with $\lambda \in \mathcal{D}_{s}$ (remember that $k$ is the material extinction coefficient responsible for optical absorption). As a consequence, even if $%
k(\lambda _0)$ is almost equal to zero, namely above the bandgap wavelength ($\lambda_g \approx 1.1 \mu$m for c-Si),
the photon, as a wave packet, can be absorbed
provided that $k(\lambda )\neq 0$ on $\mathcal{D}_{s}$. It does not mean, however, that absorption occurs below the energy bandgap ($\lambda_0 > \lambda_g$). Only available energy quanta from the wave packet ($\lambda \in \mathcal{D}_{s}$) which are above the bandgap ($\lambda<\lambda_g$) are absorbed and generate electron-hole pair.
As an illustration, let us consider a carrier wavelength equal to $\lambda_0=1700$nm for which $k(\lambda_0)\approx 0$ ($\lambda_0>\lambda_g$).
\begin{figure}
\centerline{\ \includegraphics[width=10.5 cm]{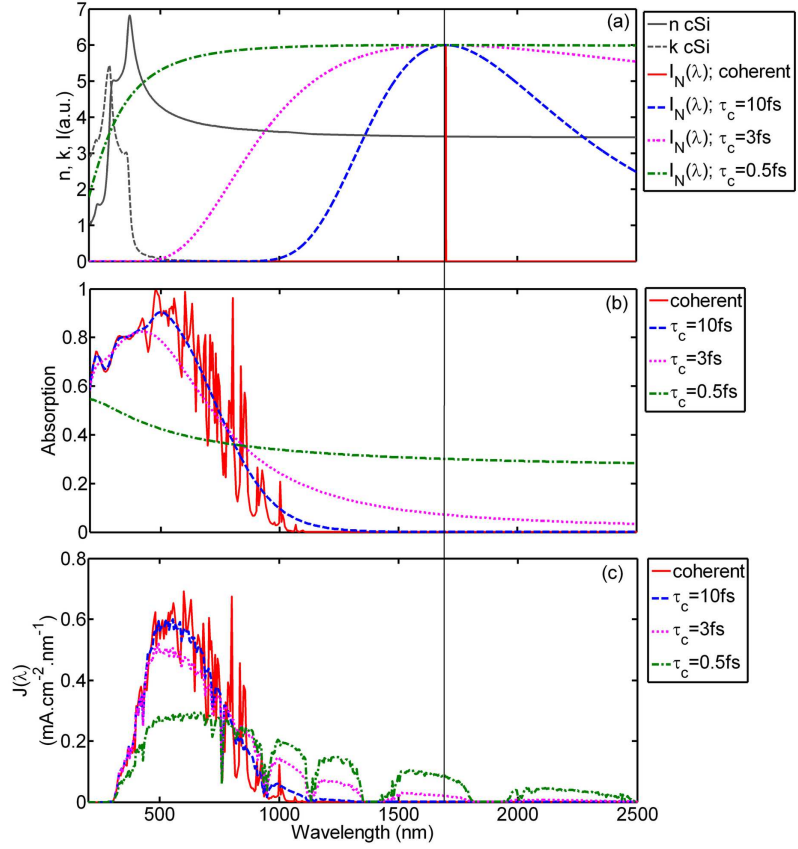}}
\caption{(Colour online). (a) Real (grey black line) and imaginary (dotted grey line) parts
of c-Si refractive index as function of wavelength. Normalized incoherence function $%
I(\lambda)$ for various coherence times: coherent case (solid red line), $\tau_c=10$fs
(dashed blue line), $3$fs (dotted magenta line) and $0.5$fs (dashed-dotted green line). $I(\lambda)$
is centered around $\lambda_0=1700$nm for illustrative purpose. (b) Absorption spectra
in a $1$ $\mu$m-thick c-Si slab with
cylindrical holes ($p=450$nm) for various coherence times: coherent case
(solid red line), $10$fs (dashed blue line), $3$fs (dotted magenta line) and $0.5$fs (dashed-dotted green
line). (c) Corresponding photocurrent spectra $J(\lambda)$ for the same values of
coherence time.}
\label{fig2}
\end{figure}

In the coherent case, when the perfectly coherent limit is reached (i.e. $\tau_c \rightarrow \infty$),
the incoherence function $I(\lambda )$
becomes a Dirac function centered
at $\lambda _0=1700$nm.
At $\lambda_0=1700$nm, the computed coherent absorption is
almost equal to zero. Accordingly,
the coherent photocurrent is also almost equal to zero at that wavelength.
In the incoherent case, the spectral width of the incoherence function $%
I(\lambda )$ increases as the coherence time decreases (figure \ref
{fig2}(a)). Therefore, a wider range of wavelengths enters into the calculation,
including shorter wavelengths that are absorbed by the material (i.e. $k\neq 0$).
Physically, as explained above, this arises from the fact that a
time-truncated sinusoidal signal ($\Delta t\approx \tau _c$), i.e. a burst of signal with finite coherence time, becomes a
polychromatic signal (with a width $\Delta E$). In other words, while the
carrier sinusoidal wave is at $\lambda _0=1700$nm, the incoherent wave packet
contains shorter wavelengths which can be absorbed. As a consequence, the whole
spectral range weightened by the incoherence function must be considered to
compute the incoherent absorption $A_{incoh}(\lambda _0)$.
This explains why the
incoherent absorption increases around $1700$nm as the coherence time
decreases.
Conversely, a wavelength (e.g. $\lambda_0=500$nm) that is strongly absorbed in the coherent
case can lead to a lower absorption in the incoherent case.
This is due to the fact that longer wavelengths experiencing
$k\approx 0$ come to play when determining the
incoherent absorption. Mathematically, the above physical considerations translate into the convolution product of (\ref{F1}).
As a result, the increase or decrease of the absorption
at a specific wavelength affects the photocurrent $J(\lambda )$ as $\tau
_c$ varies. For instance, at $\lambda _0=1700$nm, no photocurrent is
generated in the coherent case. However, as $\tau _c$ decreases, $J(\lambda _0)$ increases. On the other hand, at $%
\lambda _0=500$nm, a high photocurrent is achieved in the coherent case.
However, as $\tau _c$ decreases, $J(\lambda
_0)$ decreases. Since the total (integrated) photocurrent $J=\int
J(\lambda )d\lambda $ is obtained by integrating over a wide range of
wavelengths, it can increase or decrease according to the values of $\tau _c$ (see trends in figure \ref{fig5} below).

\begin{figure}
\centerline{\ \includegraphics[width=7.5 cm]{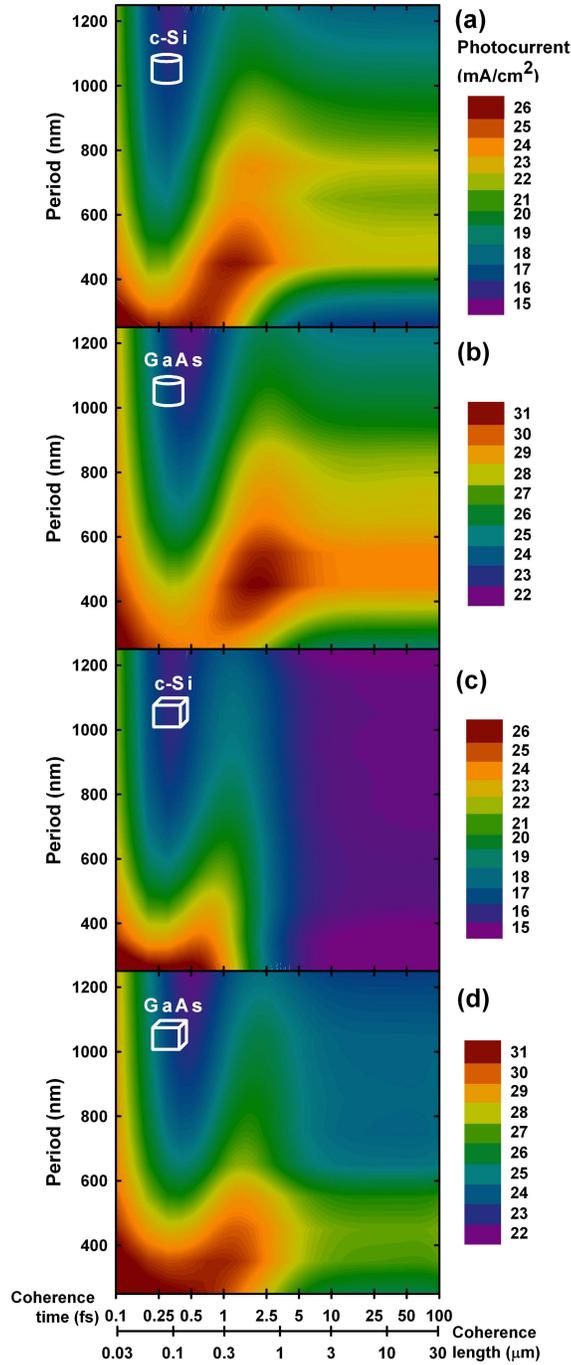}}
\caption{(Colour online). Photocurrent vs. corrugation period, coherence time or length.
Cylindrical holes in c-Si (a) or GaAs (b).
Square holes in c-Si (c) or GaAs (d).}
\label{fig3}
\end{figure}

\begin{figure}
\centerline{\ \includegraphics[width=15.0 cm]{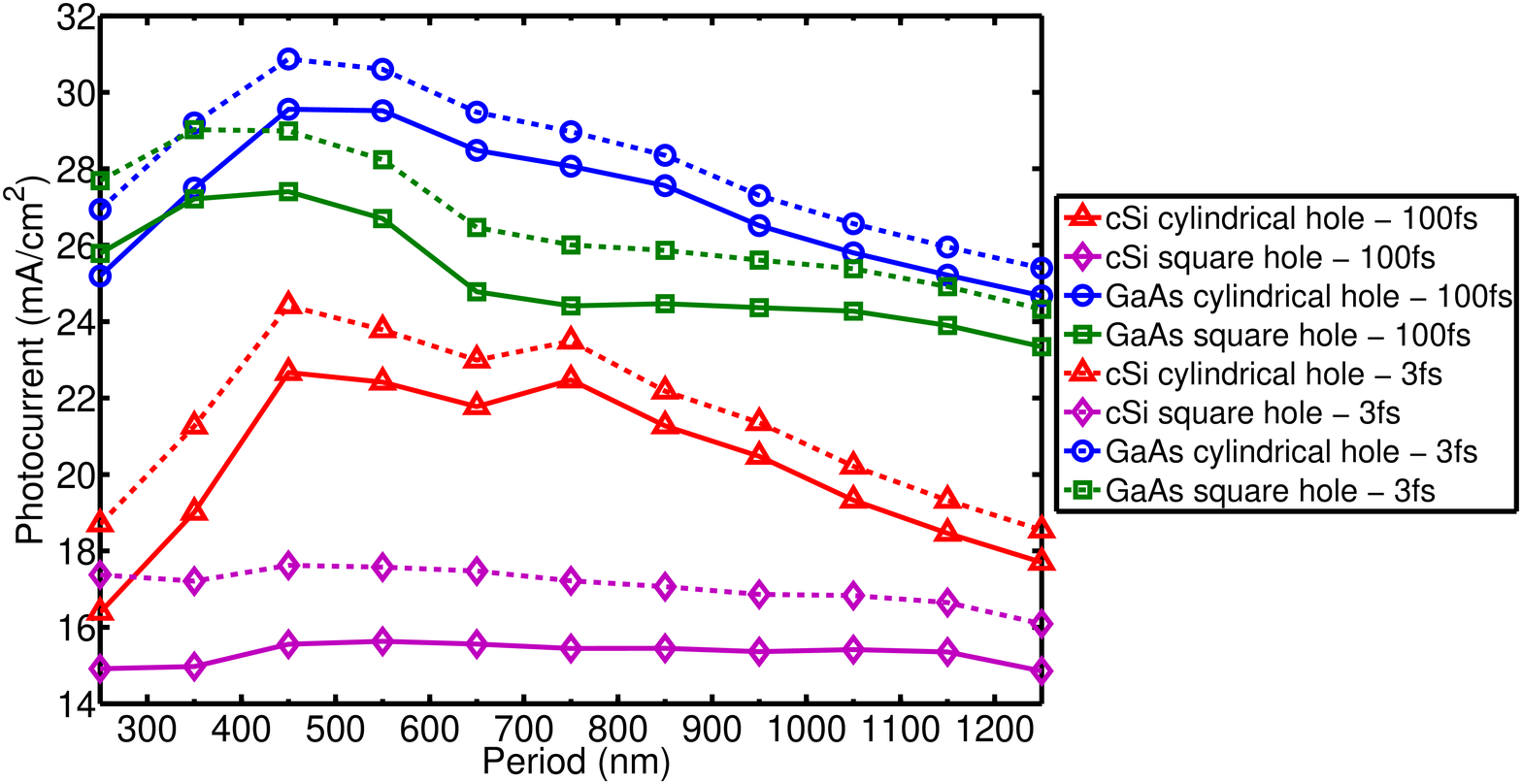}}
\caption{(Colour online). Photocurrent versus period in the (asymptotical) coherent case ($\tau_c=100$fs, solid lines) and in the incoherent case ($\tau_c=3$fs, dashed lines) for
GaAs slab corrugated with cylindrical holes (circle and blue lines) or square holes ( square and green lines)
and for c-Si slab corrugated with cylindrical holes (triangle and red lines) or
square holes (diamond and magenta lines).}
\label{fig4}
\end{figure}

Maps of the photocurrent $J$ were computed for both slabs defined in
figure \ref{fig1} with c-Si or GaAs as active material, according to various
periods, coherence times and coherence lengths (figure \ref{fig3}). 
The analysis of $J$ according to the period can be followed either in terms of coherence time ($\tau_c$) or coherence length ($l_c$). The use of $l_c$ enables the comparison between the optimal period and the estimated coherence length of sunlight.
The permittivities of materials were taken from the literature \cite{Palik}.
The aim is to highlight the effect of coherence time on the efficiency.
However, it should be noted that the relevant values in figure \ref{fig3} are those corresponding to $\tau_c$ around $3$fs (estimated coherence time of sunlight \cite{Hecht}).
In order to highlight differences between $J_{coh}$ and $J_{incoh}$, we plotted the graphs of the photocurrent ($J$) versus the period for either asymptotically coherent light ($\tau_c=100$fs) or incoherent sunlight ($\tau_c=3$fs) (figure \ref{fig4}).

Two optimal periods (i.e. maximizing the photocurrent) are found for the c-Si slab corrugated with cylindrical
holes and illuminated under coherent light: $p=450$nm and $750$nm (figure \ref{fig3}(a) and figure \ref{fig4}). If we only think in terms
of coherent light, we could use both optima since they lead to the same
photocurrent. However, when $\tau_c$ decreases, we notice that $J$
depends strongly on $\tau_c$ (figure \ref{fig3}(a)).
In a general way, we notice that, depending on the degree of coherence, a structure could be optimized
under coherent light (i.e. high values of $\tau_c$) while remaining optimal or being
better or worser under incoherent light (figure \ref{fig3}). Therefore, the choice of the optimal
corrugation (period and hole shape) strongly depends on $\tau_c$.
An optimal structure under coherent light is not
necessarily the optimal one under incoherent light, and vice versa.
For a coherence time equal to $3$fs (estimated sunlight coherence time \cite{Hecht}), in the four cases (GaAs or c-Si corrugated with square or cylindrical holes), the photocurrent is higher in the incoherent case than in the coherent case (figure \ref{fig4}). Therefore, photocurrent under sunlight could be higher than under hypothetical coherent light. This kind of behaviour has recently been observed by Abass et al. \cite{Abass_JAP}.
The choice of the optimal corrugation also depends on the
material used in the active layer.
For both materials, when the shape of the holes changes from cylinder to square, the optima shift to smaller coherence times (figure \ref{fig3}).
Since the photocurrent is strongly influenced not only by the hole shape
\cite{Herman} but also by the coherence time, it turns out to be necessary to optimize light trapping
structures taking also into account the coherence time of the solar
radiation, which is not currently done in literature.

In order to better understand the influence of $\tau_c$ on $J$, we
plotted a cross-section of the maps of figure \ref{fig3} for $p=450$nm (figure \ref{fig5}).
This period is not the optimal
one for the four studied structures. However, it is a compromise since $J$
is high for the four structures under coherent illumination. Figure \ref{fig5} shows
that the photocurrent is quite constant at long coherence times. As $\tau_c$ decreases,
however, $J$ increases, reaches a maximum, decreases and then
increases again.

\begin{figure}
\centerline{\ \includegraphics[width=12.0 cm]{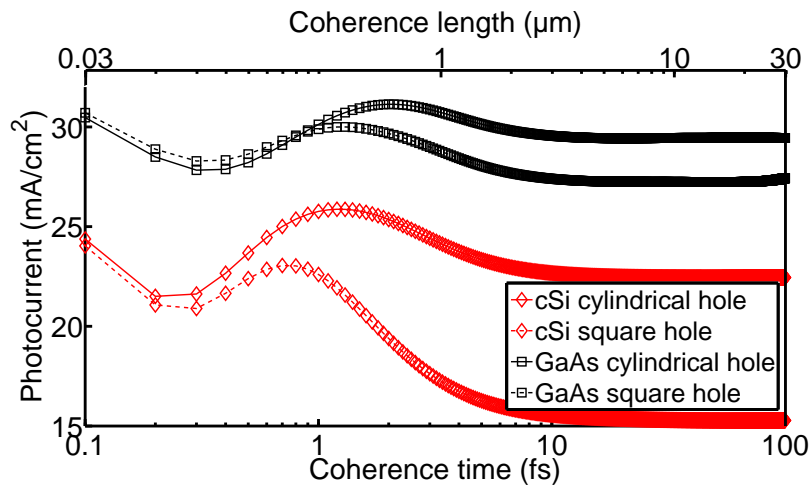}}
\caption{(Colour online). Photocurrent versus the coherence time/length for
GaAs (square and black lines) and for c-Si (diamond and red lines) slabs corrugated with cylindrical holes (solid lines) or square holes (dashed lines). 
In all cases, $p=450$nm and
$t=1 \mu$m.}
\label{fig5}
\end{figure}

Recently, researchers have investigated the effects of disorder in advanced photonic nanostructures surrounding the active layer of solar cells. These structures typically consist of complex unit cells (called super-cells) in which the nanostructure features are pseudo-randomly positioned \cite{Burresi,Pratesi,Vynck,Oskooi,Martins,Bozzola_2013,Martins_nat}. They reached the conclusion that these kinds of disordered nanostructures could increase further (in comparison with periodic nanostructures) the absorption inside the active layer. Further investigations of the impact of finite coherence time/length on disordered structures would be of high interest for future solar cell optimization.

\section{Estimation of the potential gain in computational time}

The obvious advantage of our direct method is the potential gain in computational time it offers in comparison with a multiple-run approach.
Indeed, the use of the convolution formula (\ref{F1}) in the integrated photocurrent expression (\ref{current0}) allows us to account for incoherence without the need for multiple time-consuming numerical runs and subsequent statistical analysis. In order to estimate the potential gain in computational time, let us consider as an example the $1\mu$m-thick c-Si slab corrugated with cylindrical holes. In our RCWA calculation, $11 \times 11$ plane waves (diffraction orders) were used to reach good numerical convergence. 
On our computational cluster \cite{cluster}, the calculation of the coherent absorption at a single wavelength took $t_{\lambda}=7.5$ s. The whole spectrum ($1000$ wavelengths) took therefore $t_{1000 \lambda}=7500$ s. The convolution took only a few minutes ($t_{convol}$) on a personal computer. For this calculation to be performed for $10$ grating periods (figure \ref{fig4}) took $t_{tot} \approx 10 \times t_{1000 \lambda} \approx 21$ h. If we would have used a multiple-run approach, $t_{\lambda}$ would have been multiplied by $N$, the number of runs. In Lee's method \cite{Lee}, $100$ runs were needed for each wavelength.
It would have implied a multiplication of the computational time $t_{tot}$ by a factor $N$: $t_{tot}=100 \times 21$ h $\approx 87$ days. 
Furthermore, as the complexity of the solar cell structure increases, $t_{\lambda}$ increases by orders of magnitude and therefore the total computational time $t_{tot}$ clearly becomes dissuasive using a multiple-run approach.

\section{Conclusion}
Using the theory of random signals applied to linear systems, we demonstrated that the effective incoherent absorption spectrum
of a solar cell can be directly calculated from the coherent one.
This theoretical result has a significant impact on the optimization of solar cells.
Indeed, in comparison with current numerical methods based on multiple computational runs and statistical averaging, the treatment of incoherence is shown here to no longer be related to the complexity of the cell structure, which saves a lot of computational time in many cases. The considerable simplification of the problem gives the opportunity to optimize theoretically complex solar cells under incoherent light which has been out of reach so far.

In a typical light trapping scheme based on periodic surface corrugations, we proved that the coherence time of the light illuminating
the solar cell influences drastically the maximum achievable photocurrent. Depending on
the shape of the surface corrugation and on the active layer material, the
photocurrent may increase or decrease as the coherence time changes.
In the four cases discussed in this article, the photocurrent under sunlight turns out to be higher than under coherent light.
Such a result is fundamentally related to Heisenberg uncertainty principle
and shows that solar cell efficiency may be enhanced when taking into account light incoherence. In other words, an optimal solar cell structure under
coherent illumination is not necessarily an optimal one under incoherent
illumination and vice versa. The optimization of a solar cell must therefore be performed in future by taking into account light incoherence, and not coherent 
illumination as has usually been done previously. Such a task is no longer a bottleneck since the time-consuming coherent response calculation only needs to be 
performed once for all and the incoherent response can be deduced directly from the convolution product with the power spectral density of the random process.

\section*{Acknowledgements}
The authors acknowledge J.P. Vigneron for useful discussions and comments.
M.S. is supported by the Cleanoptic project (Development of
super-hydrophobic anti-reflective coatings for solar glass panels /
Convention No.1117317) of the Greenomat program of the Wallonia Region
(Belgium). O.D. acknowledges the support of FP7 EU-project No.309127
PhotoNVoltaics (Nanophotonics for ultra-thin crystalline silicon
photovoltaics). This research used resources of the ``Plateforme Technologique de Calcul Intensif'' (PTCI)
(http://www.ptci.unamur.be) located at the University of Namur, Belgium, which is supported
by the F.R.S.-FNRS. The PTCI is member of the ``Consortium des Equipements de Calcul Intensif
(CECI)'' (http://www.ceci-hpc.be).

\end{document}